\documentclass{WileyMSP-template}

\usepackage{cite}
\usepackage[dvipsnames]{xcolor}
\usepackage{color,soul}
\usepackage{amsmath}
\usepackage{algorithm}
\usepackage{algpseudocode}
\def\bibfont{\small}
\usepackage[left]{lineno}
\usepackage{setspace}

\begin{document}

\pagestyle{fancy}
\rhead{\includegraphics[width=2.5cm]{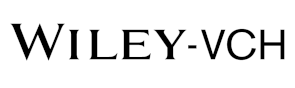}}

\title{Update Disturbance-Resilient Analog ReRAM Crossbar Arrays for In-Memory Deep Learning Accelerators}
\maketitle


\author{Wooseok Choi\textsuperscript{1*}}
\author{Tommaso Stecconi\textsuperscript{1*}}
\author{Donato Francesco Falcone\textsuperscript{1}}
\author{Matteo Galetta\textsuperscript{1}}
\author{Victoria Clerico\textsuperscript{1}}
\author{Elisa Zaccaria\textsuperscript{1}}
\author{Mamidala Saketh Ram\textsuperscript{1}}
\author{Antonio La Porta\textsuperscript{1}}
\author{Folkert Horst\textsuperscript{1}}
\author{Daniel Jubin\textsuperscript{1}}
\author{Matias Senger\textsuperscript{1}}
\author{Marilyne Sousa\textsuperscript{1}}
\author{Steffen Reidt\textsuperscript{1}}
\author{Ralph Heller\textsuperscript{1}}
\author{Bernabe Linares-Barranco\textsuperscript{2}}
\author{Valeria Bragaglia\textsuperscript{1}}
\author{Bert Jan Offrein\textsuperscript{1}}

\begin{affiliations}
\textsuperscript{1}IBM Research Europe-Zurich, 8803 Rüschlikon, Switzerland\\ 
\textsuperscript{2}Instituto de Microelectrónica de Sevilla (IMSE-CNM), CSIC and Univ. de Sevilla, 41092 Sevilla, Spain\\
E-mail: wooseok.choi@ibm.com
\textsuperscript{*}Equally contributed authors
\end{affiliations}


\begin{abstract}
Resistive memory (ReRAM) technologies with crossbar array architectures hold significant potential for analog AI accelerator hardware, enabling both in-memory inference and training. 
Recent developments have successfully demonstrated inference acceleration by \mbox{offloading} compute-heavy training workloads to off-chip digital processors.
However, in-memory acceleration of training algorithms is crucial for more sustainable and power-efficient AI, but still in an early stage of research.
This study addresses in-memory training acceleration using analog ReRAM arrays, focusing on a key challenge during fully parallel weight updates: disturbances of the weight values in cross-point devices. A ReRAM device solution is presented on 350 nm silicon technology, utilizing a resistive switching conductive metal oxide (CMO) formed on a nanoscale conductive filament within a HfO$_{\rm x}$ layer.
The devices not only exhibit 60 ns fast, non-volatile analog switching, but also demonstrates outstanding resilience to update disturbances, enduring over 100k pulses.
The disturbance tolerance of the ReRAM is analyzed using COMSOL Multiphysics simulations, modeling the filament-induced thermoelectric energy concentration that results in a highly nonlinear device responses to input voltage amplitudes.
Disturbance-free parallel weight mapping is also demonstrated on the back-end-of-line integrated ReRAM array chip. 
Finally, comprehensive hardware-aware neural network simulations validate the potential of our ReRAM for in-memory deep learning accelerators capable of fully parallel weight updates.

\end{abstract}

\section{Introduction}
\quad In the era of artificial intelligence (AI), emerging memory technologies such as resistive memory (ReRAM) have garnered significant attention for neuromorphic devices due to their high scalability, low-power operation, and analog resistive switching capability\textsuperscript{\cite{mannocci2023rram, wu2018methodology, mehonic2024rram, boybat2024pcm, burr2017neuromorphic, huang2024memristor,falcone2024nanoscale,clerico2025edge}}. Embedded in a crossbar architecture, analog in-memory computing has been successful in accelerating AI inference, enabling fully parallel dot product operations\textsuperscript{\cite{hu2018AM, narayanan2021ted, wan2022nature, song2024science, manueal2023natureelec, choi2021hardware,choi2020TNANO}}. Advanced inference chips are anticipated to show $\times140$ higher energy efficiency compared to digital systems\textsuperscript{\cite{jain2022tvlsi}}. As they do not support the computationally expensive training, the training workloads are outsourced to off-chip digital processors.

\quad Meanwhile, the rapid advancement of AI algorithms is pushing digital hardware systems to their limits in handling training costs\textsuperscript{\cite{AI2024report, cottier2024ai, gholami2024aimemorywall}}. The size of neural networks (NNs) has grown at an unprecedented rate, surpassing trillions of parameters. 
As a result, the bottleneck problem inherent in digital computing architectures has led to constraints in power efficiency, causing AI training costs to exceed 100 million US dollars for state-of-the-art models\textsuperscript{\cite{AI2024report, cottier2024ai, gholami2024aimemorywall}}. 
Thus, an in-memory deep learning accelerator capable of fully parallel weight updates offers a path to more sustainable AI. 

\quad {There have been numerous simulation studies on how device properties affect in-memory on-chip training\textsuperscript{\cite{chen2018neurosim, le2023aihwkit, JWJang2015EDL, Burr2015TED, xiao2022crosssim}}.} However, experimental studies utilizing analog memory arrays still remain in an early stage (Figure S1)\textsuperscript{\cite{aguirre2024frontiers,gong2022iedm,li2021frontiers,kim2022frontiers,fuller2019science,kim2019ud,gao2015nanotech, manueal2023natureelec,wan2022nature,narayanan2021ted,hung2021NatureElec,song2024science,wu2023AM,hu2018AM,yao2020nature}}. {To enable scaling up the AI training accelerator system with larger matrix arrays, systematic studies combining experimental and simulation work—while considering practical parallel array update schemes—are essential for bridging this gap. 
In 2016,} Gokmen et al. introduced a key challenge, that is update disturbances in cross-point devices during parallel array updates through simulation work\textsuperscript{\cite{gokmen2016acceleration}}.
{Although update disturbances have been extensively studied in digital memory applications, conventional standards for evaluating these disturbances in digital memories are not directly applicable to analog memory devices and in-memory training systems. For example, in storage-class memory systems, a disturbance of up to a few percent in the conductance state can still be acceptable for reading discrete multiple bits. In contrast, even small disturbances in analog synaptic memory can significantly degrade learning accuracy in AI hardware by hindering the convergence of granular memory states to their desired values, especially when the disturbances accumulate during continuous learning.
Despite their importance, studies on update disturbances remain limited, with most existing results focusing on single devices\textsuperscript{\cite{kim2019ud, kim2023ud, kim2024ud, chung2024ud}}. In this context, extensive studies presenting thorough experimental data on update disturbances and their system-level evaluations are essential.}

\quad In this work, we demonstrate a disturbance-resilient analog ReRAM chip developed on 350 nm silicon technology, using CMOS-compatible conductive metal oxide (CMO)/HfO$_{\rm x}$ materials. We rigorously verify disturbance resilience in one-transistor one-ReRAM (1T1R) cells and arrays, considering realistic in-memory update scenarios. COMSOL Multiphysics simulations correlate the disturbance-resilient device properties with resistive switching physics, revealing that the filament-induced thermoelectric energy localization leads to a highly nonlinear switching responses to the input voltage amplitudes. Furthermore, we showcase disturbance-free parallel weight mapping on a wire-bonded 1T1R array. 
Finally, we present systematic NN simulations that incorporates all relevant ReRAM properties and the parallel update scheme, proving the potential of our novel ReRAM for fully in-memory AI accelerators. 


\section{Background}
\subsection{In-Memory Outer-Product Weight Update}
\quad To train a deep NN, the weight gradients need to be calculated for every iteration by computing the outer-product of the forward activations $\textbf{x}$ and the backpropagated errors $\textbf{d}$. Then, the gradients are accumulated to the current weight values ($W \leftarrow W + \eta \textbf{x} \textbf{d}^T$, where $\eta$ is a learning rate)\textsuperscript{\cite{choi2021neural}}.
For in-memory training accelerations, this outer-product weight update must be performed in-memory and parallel across the analog memory arrays\textsuperscript{\cite{rasch2024NatureComm}}. This section describes the principle of the in-memory outer-product update on the array hardware and unwanted disturbances in cross-point devices. 
\begin{figure}[!t]
    \begin{center}
      \includegraphics[width=\linewidth]{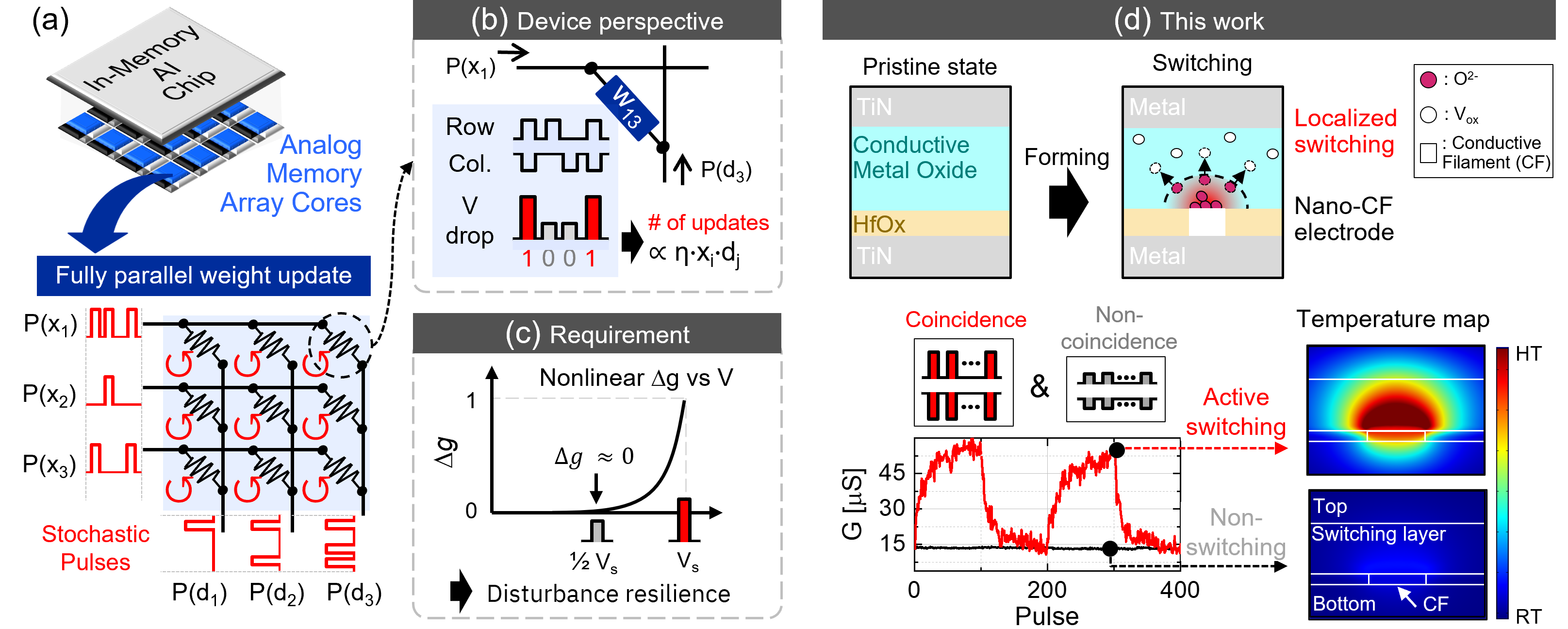}
      \caption{
      (a) An analog AI accelerator and the fully parallel weight update of one array core. 
      (b) Illustrations of a stochastic parallel weight update scheme. In a single device perspective located in i$_{\rm th}$ row and j$_{\rm th}$ column, the number of coinciding pulses becomes proportional to the nominal amount of weight update $\eta x_i d_j$.
      (c) To prevent undesired weight changes during simultaneous array updating, the device should have strongly nonlinear responses to the applied voltage amplitudes.
      (d) This study develops a filamentary analog ReRAM that is high resilient to the update disturbance, enabling the outer-product weight updates fully in-memory.}
      \label{fig:F1}
    \end{center}
\end{figure}

\quad Here, a stochastic pulse encoding scheme reported by Gokmen et al. enables the fully parallel outer-product update in constant time complexity $\mathcal{O}(1)$\textsuperscript{\cite{gokmen2016acceleration, haensch2018IEEEproceedings}} (\textbf{Figure 1a}). The core of this scheme is the probabilistic coincidence of voltage pulses, to which cross-point devices respond under a full switching voltage $V_s$. When the stochastic pulse trains with half the amplitude of $V_s$ are submitted to all array inputs simultaneously, the number of coinciding pulses in a weight matrix becomes proportional to the outer-product of $x$ and $d$. \textbf{Figure 1b} depicts a device perspective located at i$_{\rm th}$ row and j$_{\rm th}$ column in the array.
\textbf{Equation \ref{E}} shows an example of the probability-encoded activations, $P(\textbf{x})$ and $P(\textbf{d})$, and the resulting outer-product $P(\textbf{x})P(\textbf{d})^T$ ($\propto \textbf{x} \textbf{d}^T$).
\begin{equation} \label{E}
    P(\textbf{x}) = 
    \begin{pmatrix}
        p(x_1) \\ p(x_2) \\ p(x_3)
    \end{pmatrix}
    , \quad
    P(\textbf{d}) = 
    \begin{pmatrix}
        p(d_1) \\ p(d_2) \\ p(d_3)
    \end{pmatrix}
    , \quad
    P(\textbf{x})P(\textbf{d})^T = 
    \begin{pmatrix}
        p(x_1)p(d_1) & p(x_1)p(d_2) & p(x_1)p(d_3) \\
        p(x_2)p(d_1) & p(x_2)p(d_2) & p(x_2)p(d_3) \\
        p(x_3)p(d_1) & p(x_3)p(d_2) & p(x_3)p(d_3) \\
    \end{pmatrix}
\end{equation}

The main advantage of this scheme is its ability to perform both gradient calculations and updates by simply providing stochastic pulse trains to the array inputs, as the resistive devices automatically change their resistance states by detecting the pulse coincidence with full $V_s$ amplitude. This method indeed holds significant potential for in-memory deep learning accelerators, since the operational time $\mathcal{O}(1)$ is independent to the size of weight matrices.

\quad For successful training, however, the cross-point devices must ignore non-coincident half $V_s$ pulses while responding only to coinciding pulses. Since stochastic pulse trains of $p(x_i)$ and $p(d_j)$ are sparse in real applications, the number of non-coincident pulses exceeds the number of coincident ones (Figure S2). Therefore, a strong non-linearity in resistive switching in response to applied voltage amplitudes is highly desirable to avoid update disturbances (\textbf{Figure 1c}). Previous studies have employed multiple transistors in each cross-point element to filter out the coinciding pulses from non-coincident pulses\textsuperscript{\cite{Kohda2020IEDM, gong2022iedm, won2023AdvSci}}. However, it incurs a significant penalty in memory density and system scalability. In the following sections, we present our device solution with inherently disturbance-tolerant, analog ReRAM technology by enhancing highly localized switching activations thanks to the pre-formed conductive filament (\textbf{Figure 1d}).


\section{Results}
\subsection{Conductive Metal Oxide/HfO$_{\rm x}$ Analog ReRAM}
\begin{figure}[!t]
  \includegraphics[width=\linewidth]{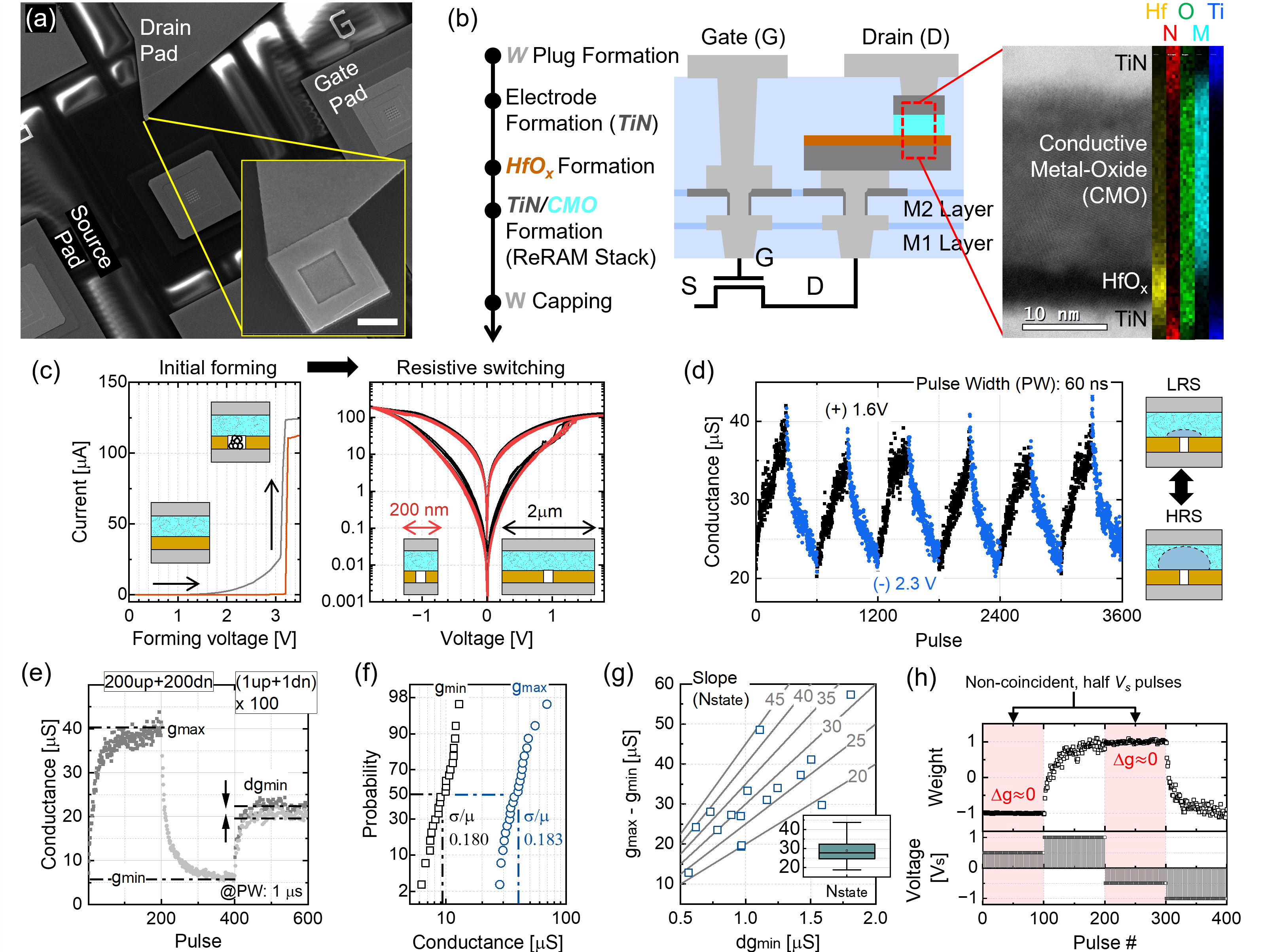}
  \caption{
  (a) SEM image of a 1T1R unit cell and the effective device area. The scale bar represents 1\textmu m. 
  (b) Integration process flow of ReRAM with 350 nm silicon technolgy. The TEM image and EDS elemental mapping profile demonstrate the device material structure. 
  (c) Initial filament forming and the following resistive switching of ReRAMs (device dimension: 200 nm and 2 \textmu m). 
  (d) 60 ns pulsed cycling with +1.6 V/-2.3 V amplitudes. 
  (e) Extracting the key device metrics from the pulse measurements with 1 \textmu s and +1.4 V/-1.9 V.
  (f) Probability plot of $g_{\rm min}$ and $g_{\rm max}$ distributions.
  (g) Scatter plot showing the conductance range and $dg_{\rm min}$, where the slope represents the effective number of states ($N_{\rm state}$) during open-loop updating. Each data point represent a different individual device.
  (h) Update disturbance test results and the used pulse scheme. The red shaded area indicates when the device gets non-coincident, half $V_s$ pulses.}
  \label{fig:F2}
\end{figure}

\quad \textbf{Figure \ref{fig:F2}a} shows scanning electron microscope (SEM) images of the 1T1R cell integrated with 350 nm silicon technology. The highlighted inset indicates the active device area of ReRAM. The Back-End-Of-Line (BEOL) integration process flow of the ReRAM is described in \textbf{Figure \ref{fig:F2}b}. The scanning transmission electron microscopy (STEM) image shows a device cross-section consisting of a conductive metal oxide (CMO)/HfO$_{\rm x}$ bilayer structure with TiN electrodes. Through energy dispersive X-ray spectroscopy (EDS) analysis, the elemental mapping profile verifies the material structure of the device. {To facilitate further exploration and innovation in CMO/HfO$_{\rm x}$ analog ReRAM technology, a list of material candidates for the CMO layer is provided in Table S1.}

\quad At the pristine state, a conductive filament (CF) in the insulating HfO$_{\rm x}$ layer needs to be formed by applying a controlled voltage and current to the device. \textbf{Figure 2c} shows the forming of 1T1R cells as a function of applied bias. The single-step and abrupt current jump demonstrates the formation of a rigid CF by the breakdown through the 4 nm HfO$_{\rm x}$ film. The effective forming voltage across the device was found to be 2.8 V on average (Figure S3). The right graph in \textbf{Figure 2c} depicts the gradual set and reset switching upon applying positive and negative voltage sweeps. The area-independent operations also confirm the role of the CF in defining the confined resistive switching region within the resistive switching CMO layer. {Our previous study demonstrated that, with the rigid CF, electronic transport in the device is governed by trap-to-trap tunneling through the CMO layer in both the low and high resistance states\textsuperscript{\cite{galetta2024compact,falcone2024nanoscale,stecconi2024analog,falcone2025all}}. Based on these observations, the resistive switching mechanism was explained by electric field and temperature induced modulation of trap density within the sub-band of the CMO layer\textsuperscript{\cite{falcone2025all}}. Thus, resistance changes in the ReRAM device are attributed to the migration of oxygen ions (or defects) within the CMO layer, specifically the half-spherical volume at the interface with the CF\textsuperscript{\cite{falcone2025all}}. \textbf{Figure 2d} presents 60 ns fast pulse measurements, applying +1.6 V for conductance increase (set process) and -2.3 V for conductance decrease (reset process). Negative pulses induce defect migration away from the half-spherical volume above the CF, oxidizing the region and increasing the device resistance. Conversely, positive pulses promote the back-migration of defects toward the volume, restoring a lower-resistance state. The right insets in \textbf{Figure 2d} illustrate the change in the effective volume with a low trap density in the CMO due to defect migration.}

\quad \textbf{Figure 2e} presents the experiments for extracting key device parameters for deep learning applications\textsuperscript{\cite{rasch2021AICAS, stecconi2024analog}}. The large energy per pulse with a longer duration drives the device conductance into saturation, both at the high- and low-conductance extremes. {Note that the differences in linearity and analog modulation observed in Figures 2d and 2e can be attributed to the different programming conditions used in each experiment\textsuperscript{\cite{frascaroli2018softbounds,jacobs2017linearity}}.} Although we have confirmed the tens of nanoseconds switching capabilities of our devices, we adopt a microsecond-scale pulse width for the following experiments on a wire-bonded devices by using our array controller setup.
The two distinct distributions of maximum and minimum conductance ($g_{\rm max}$ and $g_{\rm min}$) are shown in the \textbf{Figure 2f}, obtained from 20 devices. According to the soft-bounds model for analog memory devices\textsuperscript{\cite{gong2022iedm, stecconi2024analog,rasch2024NatureComm}}, the effective number of states $N_{\rm state}$ is defined as the ratio of the operating conductance range to the minimum G change at the balanced symmetry point (SP) $dg_{\rm min}$ (The fitted soft-bounds model of our device is presented in Figure S4). 
Thereby, the slope in \textbf{Figure 2g} represents the $N_{\rm state}$ of each analog ReRAM device. The inset box plot shows an average $N_{\rm state}$ of 27 conductance levels. {Note that the $N_{\rm state}$ parameter quantifies the degree of analog modulation that the synaptic memory can achieve during on-chip learning with open-loop updating\textsuperscript{\cite{rasch2024NatureComm}}. Thus, it does not indicate 27 distinct, non-overlapping multilevel states.}
Here, we reveal update disturbance test results considering the pulse scheme in stochastic parallel array update, introduced in Figure 1a. \textbf{Figure 2h} shows the experimental weight changes of our ReRAM device by consecutively applying 4 different pulse trains to the device; 100 non-coincident up, 100 coincident up, 100 non-coincident down, and 100 coincident down pulses. The electrical measurements were conducted on a wire-bonded 1T1R cell by using pulses with +1.4 V/-1.9 V and 2.5 \textmu s. 
{It is important to note that the disturbance was tested in the worst case scenario at each conductance bound (i.e., $g_{\rm min}$ and $g_{\rm max}$), where the device has the largest momentum to change its conductance towards the other boundary (can also be seen in Figure S4). By giving more consideration to the steady-state relaxation of the material system, rather than the dynamic on-chip training scenario, the evaluation of disturbances can also be performed at different states.} Nevertheless, our analog ReRAM technology demonstrates its superior resilience to the half $V_s$ pulses, i.e., non-coincidence cases in stochastic fully parallel array update. {The retention of 32 intermediate conductance states over 100 seconds was evaluated to demonstrate non-volatility, including short-term relaxation effects (Figure S5a). The cycling endurance of our ReRAM device, exceeding 100 million cycles, was also demonstrated in Figure S5b.}

\subsection{ReRAM Model}
\quad To better understand the device physics, we conducted 3D finite element simulations in COMSOL Multiphysics by solving the continuity and Joule-heating equations in steady state (Supplementary Note 1). The electric field and temperature distributions within the switching layers were extracted, considering the experimental $IV$ data using the same approach as reported by our previous works\textsuperscript{\cite{falcone2024nanoscale, galetta2024compact}}. \textbf{Figure 3a} shows the $IV$ response of the device from measurements (gray line) and fitted model (dashed red line). The model reproduces the $IV$ response at the highest resistance state (HRS) and lowest resistance state (LRS) along with the points A and B, from which the device begins resistive switching in both polarities. The material parameters used for the simulations are described in the graph. As noted in the Figure 3a, the radius of the conductive filament (white), $r_{\rm CF}$, in the insulating HfO$_{\rm x}$ layer (brown) was estimated to be approximately 11 nm by modeling the electronic conduction in the low-voltage linear regime of the LRS. In \textbf{Figure 3b}, the 2D color plot on the xy plane of the graph shows that the CF acts as a bottom electrode extension, thereby most of the electric field drops at the interface between the CMO and CF. This causes a highly concentrated electric field in the active switching area in the CMO layer. Consequently, the electric field-induced current leads to a strong temperature increase in the confined region within the CMO layer. Thereby, \textbf{Figure 3c} shows the highly localized heat map in the device structure at points A and B (also indicated in Figure 3a). The temperature concentration can be clearly observed above the nanoscale CF in the simulation results. 
\begin{figure}[t]
  \includegraphics[width=\linewidth]{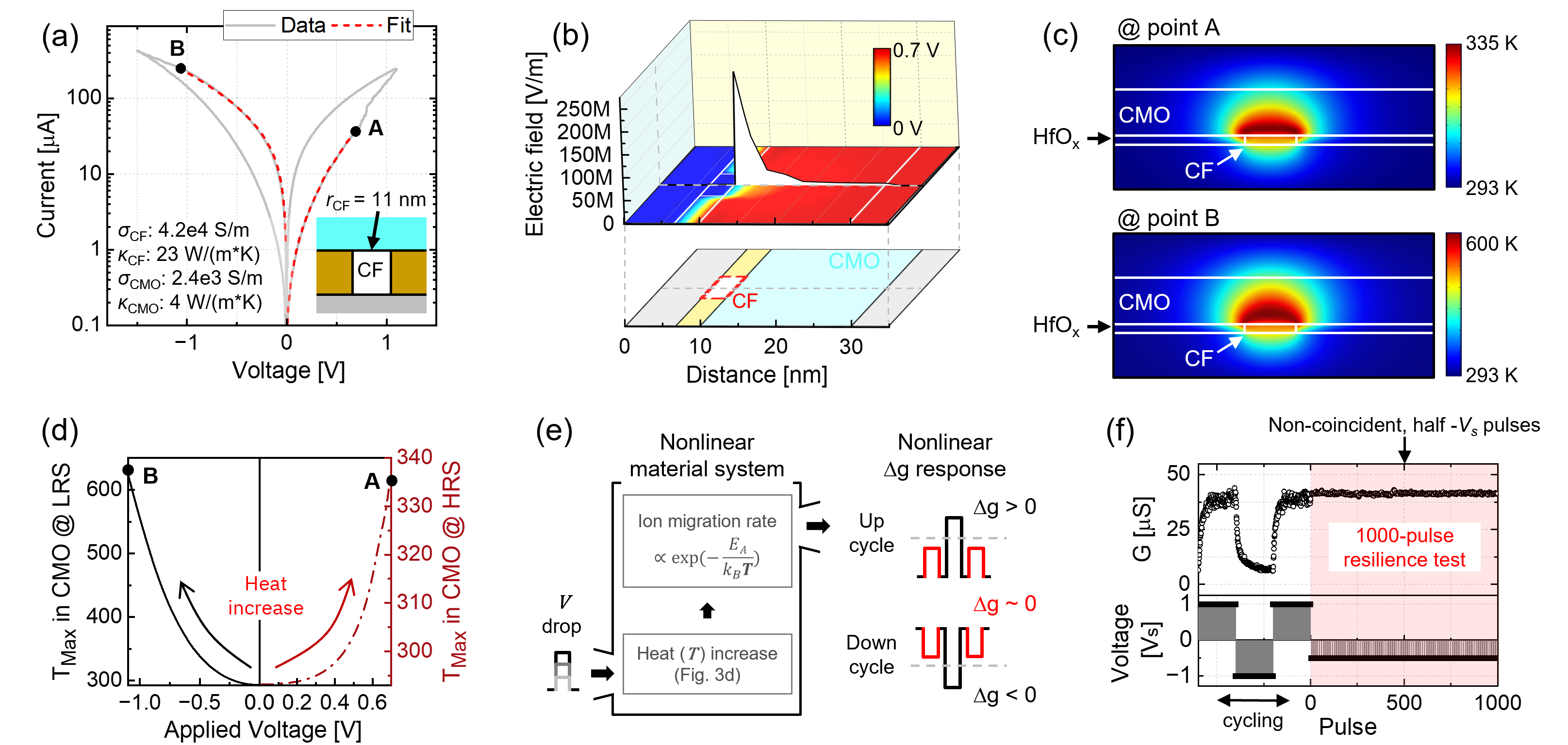}
    \caption{
    (a)$IV$ curve of the ReRAM (gray) and the fitted lines (red) from COMSOL simulations. The endpoints, A and B, on each fitted line are marked at +0.7 V and -1.1 V, respectively.
    (b) 2D color plot on the $xy$-plane showing the electric potential map at Point A, aligned with the device structure illustrated below from the bottom (left) to the top electrodes (right). The conductive filament (CF) is indicated by a red dashed square. The graph presents the electric field line profile across the device demonstrating the extreme concentration of electric field on top of the electrode-like CF.
    (c) Highly localized temperature profiles around the nanoscale CF in the CMO resistive switching material at Points A and B.
    (d) Simulated maximum temperature $T_{\rm Max}$ across the 3D volume of the CMO layer. The applied voltage significantly increases device internal temperatures.
    (e) Schematic of bias-induced resistive switching process, resulting in a highly nonlinear $\Delta$g response. The applied electric potential not only lowers the activation energy $E_A$, but also increases the device's internal temperature due to Joule heating, thereby facilitating the ion migration rate exponentially.
    (f) Reliability test of update disturbanceunder 1000 pulses in the lowest resistance state.}
  \label{fig:F3}
\end{figure}

\quad We further investigate a maximum temperature $T_{\rm Max}$ in the CMO film as a function of the applied voltage amplitudes. In \textbf{Figure 3d}, $T_{\rm Max}$ in the resistive switching CMO layer significantly increases with the applied voltage amplitude. The dash-dot line represents $T_{\rm Max}$ at HRS during the voltage sweep in the set direction and a solid line represents $T_{\rm Max}$ at LRS for the reset direction. The points A and B are also noted here at the ends of each line, where the device starts the set and reset switching processes, respectively.
 From these points, even a small increase in voltage can accelerate current-temperature feedback, facilitating resistive switching. Thereby, the conductance change $\Delta$ g becomes highly nonlinear to the applied pulse amplitude such that the device does not change its conductance at a half $V_s$ pulses. 
This inference can also be supported by the Arrhenius equation of ion migration rate, which is significantly accelerated by the local temperature increase (Supplementary note 1)\textsuperscript{\cite{lelmini2011TED, padovani2015TED}}.
The nonlinear material system is summarized in \textbf{Figure 3e}. The voltage pulse causes the nonlinear heat $T$ increase, facilitating the ion migration rate exponentially\textsuperscript{\cite{lelmini2011TED, padovani2015TED}}. 
The activation energy $E_A$ barrier is also lowered  by the magnitude of applied electric potential.
As a result, the nanoscale CF-based confined switching activation of our ReRAM enhances the non-linearity in resistive switching with respect to the input voltage amplitude.
It is worth noting that a further scaled CF (less than 10 nm radius) can localize thermoelectric energy more effectively, which potentially enhance the switching non-linearity with respect to the applied voltage amplitude (Figure S6). 
We also studied an extreme case when the device faces 1000 non-coincident pulses at the lowest resistance state (\textbf{Figure \ref{fig:F3}f}). We regard the LRS disruption as the worst case scenario, where the device has a high thermal state under the half $V_s$ pulse voltage (also can be seen in Figure 3d). The stable resistance state even under 1000 half $V_s$ pulses demonstrate remarkable disturbance resilience of our filamentary analog ReRAM technology.

\subsection{Analog ReRAM Crossbar Array}
\quad As shown in \textbf{Figure 4a}, we performed electrical characterizations of a 5$\times$5 array by using a custom-built testing system capable of generating arbitrary waveform signals and simultaneously measuring the output responses on the multiple channels individually (Figure S7). \textbf{Figure 4b} shows the single memory access scheme in the 3$\times$3 subarray by submitting half $V_s$ pulses to the corresponding row and column such that the coincident pulses with full $V_s$ happen at the target cell. In the inset graph in Figure 4b, we observe that the conductance states of the devices were not disturbed by the hundreds of half $V_s$ pulses. {Figure S8 shows the average conductance values with one-sigma error bars obtained from 100-pulse cycling experiments conducted on multiple devices within the array.}
\begin{figure}[t] \includegraphics[width=\linewidth]{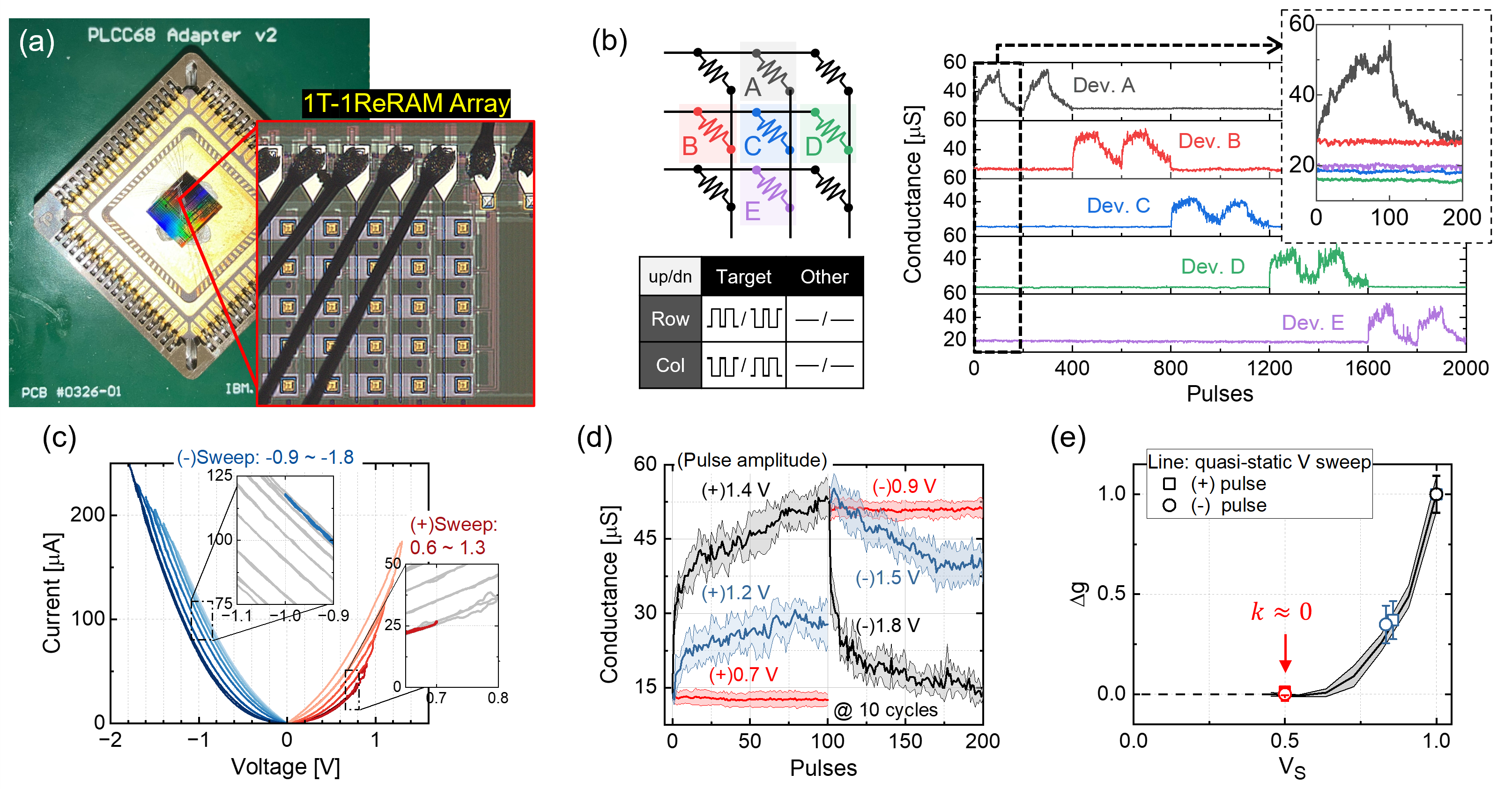}
    \caption{
    (a) Printed circuit board (PCB) with a wire-bonded, 5$\times$5 1T1R array chip used in the experiments.
    (b) Single memory access in the analog ReRAM array using a half $V_s$ pulse scheme. The analog memory states of the target device are gradually updated, while the states of other cells remain unaffected by non-coincident pulses.
    (c) Quasi-static voltage sweep measurements. The device is resistant to conductance changes even with the quasi-static voltage bias higher than half $V_s$ on both switching direction.
    (d) Device resistive switching responses to a set of 100 pulses with various voltage amplitudes, which are mapped by colors in the graph. The data was obtained from 10 cycles by using 2.5 \textmu s pulse width.
    (e) The CMO/HfO$_{\rm x}$ ReRAM demonstrates the non-linearity factor $k$ close to zero, indicating negligible half $V_s$ disturbance.}
  \label{fig:F4}
\end{figure}
We also conducted stability tests of the resistance states under quasi-static voltage sweeps and pulses. As shown by the red lines in \textbf{Figure 4c}, positive voltage sweeps were applied to the highly resistive device where the peak values ranged from 0.6 V to 1.3 V with a 0.1 V step increase. The inset graph clearly shows that the device was not affected by the demanding 0.7 V stress signal. Likewise, the measurements on the negative polarity confirmed the robust resistance state even under -1 V stress (blue lines). The pulse measurement results were also obtained by using 2.5 \textmu s fast pulse width as shown in \textbf{Figure 4d} with different colors depending on the used pulse amplitudes. The shaded area displays one-sigma distributions across multiple measurement cycles. The black line shows the baseline switching progress with full $V_s$, whereas the red line indicates the device response to half $V_s$ pulses. As defined in the previous study, the non-linearity factor $k$ in \textbf{Equation \ref{E1}} represents the ratio of two conductance change results ($\Delta$ g) at half and full $V_s$ amplitudes.
\begin{equation} \label{E1}
    k = \frac{ \Delta g(0.5V_s)}{\Delta g(V_s)}
\end{equation}
where the ideal $k$ value is 0. The experimentally obtained $k$ value of our devices is close to zero, as shown in \textbf{Figure 4e}. 
The black solid line with shaded area represents the results of quasi-static measurements across the voltage with one-sigma variation. Also, the results of pulse measurements (symbols) correspond well with those of quasi-static measurements. The colors of the symbols indicate the pulse scheme noted in Figure 4d and the symbol shapes represent the update directions in the experiments. This high non-linearity ensures that the devices are intrinsically robust to the update-disturbances, enabling fully parallel weight updates in a highly scalable ReRAM array architecture. Note that the $\Delta$ g on the y-axis and $V_s$ on the x-axis were normalized to the conductance range and the operating voltage used in the experiments.

\quad {We extended our evaluation of the update disturbance test to include up to 1 million (1M) non-coincident pulses at each conductance bound across multiple devices in the array (\textbf{Figure 5a}). For the test, the polarity of the pulses is chosen such that they drive the device conductance (G) away from high-G and low-G set values. The effective voltage drops of the non-coincident pulses were +0.7 V and –0.9 V, with a pulse width of 2.5 \textmu s. The results experimentally demonstrate the strong update resilience of our analog ReRAM technology. Moreover, we analyzed the magnitude of conductance changes as a function of the number of update disturbance pulses (\textbf{Figure 5b}), and the G change is shown in \textbf{Figure 5c}. Under this harsh condition of applying 1M disturbance pulses, the device G does not show any tendency to drift even after 100k pulses (Figure S9). Note that the high G states are more susceptible to Joule heating effects compared to the low G states. Thereby, the high G states start showing downward disturbances after receiving 1 M non-coincident pulses. Note that this test was conducted by assuming extreme worst-case scenarios, where devices continuously experience 1M non-coincident pulses solely at each G bound during on-chip training. Hence, we are convinced that the observation of stable conductance states up to 100k disturbance pulses provides a demonstration of update disturbance resilience.}

\quad The scatter plot in \textbf{Figure 5d} displays the specified $k$ values (i.e., $k_{\rm up}$ and $k_{\rm dn}$) with hundred data points collected from 20 devices, including device-to-device and cycle-to-cycle variations. The black and red symbols represent the non-linearity factors after DC voltage sweeps up to the corresponding voltage amplitudes and after 100 disturbance pulse tests, respectively. The non-linearity factors after 100k disturbance pulse tests are also indicated by the green symbols from 15 devices. We observe the variations in $k$ values reaching up to 0.05, which result from conductance fluctuations under voltage pulses, caused by metastable oxygen ion movements within the CMO layer (Figure S9). The mean ($\mu$) and standard deviation ($\sigma$) of both $k_{\rm dn}$ and $k_{\rm up}$ after 100k pulses are presented in Figure 5d by the green box plots and the table. The $\mu$ and $\sigma$ were less than 0.005 and 0.025, respectively. Here, we describe an effective non-linearity factor of our device technology as the mean of the $k$ distribution, which is less than 0.005. This interpretation of $k$ allows us to effectively present the $k$ value of our ReRAM technology in the presence of G fluctuations. Note that the averaging will only cancel out the variations, not the disturbance-induced G drifts. If the devices show any tendency of G drift away from the initial state, the effective k value will become significant, even after averaging. It is also worth noting that the results at 100 pulses in Figure 5c correspond well with those shown in Figure 5d, validating the consistency of our test results.
\begin{figure}[!t]
  \includegraphics[width=\linewidth]{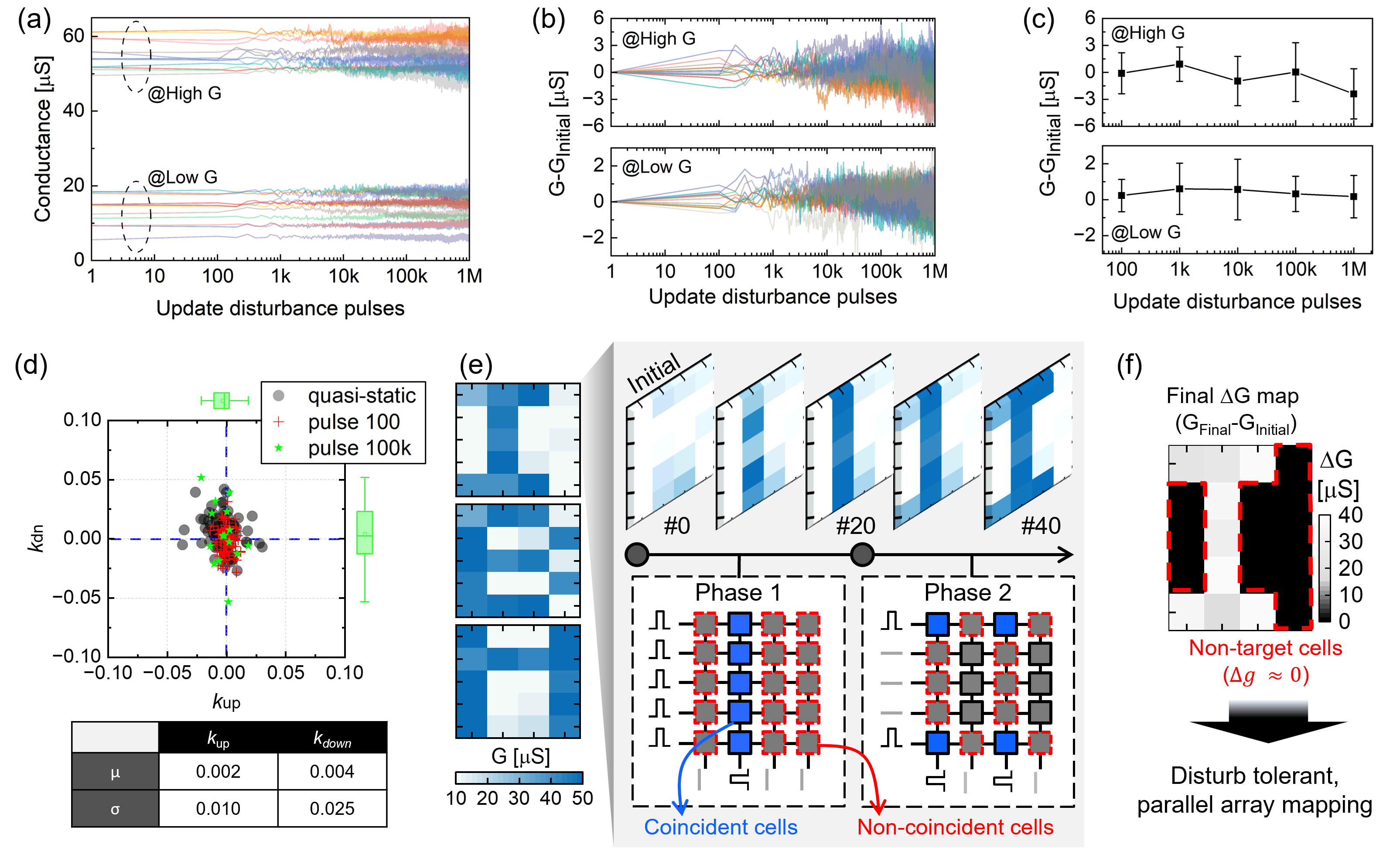}
    \caption{{(a) Update disturbance tests under extreme conditions, applying 1 million non-coincident pulses to each conductance boundary across multiple devices in the array. The devices were programmed into their respective conductance bounds before undergoing the non-coincident pulses. The effective voltage drops of the non-coincident pulses were +0.7 V and –0.9 V, with a pulse width of 2.5 \textmu s. (b) Conductance difference from the initial state during the 1 million pulse tests at both high and low boundary states. (c) Conductance changes relative to the operating conductance window as a function of update disturbance pulses.}  
    {(d) Scatter plot showing the non-linearity factors for up ($k_{\rm up}$) and down ($k_{\rm dn}$) directions from various measurements. The box plots demonstrate that the distribution of $k$ values after 100k pulse tests centered around zero. The table presents the mean ($\mu$) and standard deviation ($\sigma$) of both $k_{\rm dn}$ and $k_{\rm up}$ after 100k pulses.
    (e) Letter mapping demonstrations on a 5 × 4 array. The update process is shown in the shaded area. Half $V_s$ pulses are applied to the array rows and columns such that coinciding pulses with a full $V_s$ drop arrive at the target cross-point devices. The stochastic pulses with a probability of 0.5 and bitstream lengths ($BL$) of 10 were used in the experiments.  
    (f) The final grayscale pseudocolor plot of $\Delta$g confirms the device's update disturbance tolerance during parallel crossbar updates.}
    }
  \label{fig:F5}
\end{figure}

\quad Additionally, we demonstrate disturbance-free, parallel array mapping on the ReRAM array chip. \textbf{Figure 5e} shows a representative case of the “I” letter with the evolution of the array conductance map. The mapping scheme is also displayed in the illustrations for phase 1 and 2. 
Note that the experiments were conducted by submitting stochastic pulse trains to all array inputs where the target cells are located. A bitstream length ($BL$) of 10 and a probability of 0.5 were used for the stochastic pulse trains. Thereby, the scheme reflects a similar situation to the stochastic fully parallel weight update scenario in NN training (Figure S10).
Note that the digital letters are mapped throughout enough iterations and $BL$, which allows us to establish extreme conditions for testing update disturbance during parallel crossbar updates. The extension of this study will involve NN training assessments that observe analog weight learning. 
The experimentally generated stochastic pulse trains and the measured (non-)coincidence examples are also shown in Figure S11. \textbf{Figure 5f} shows the final conductance changes, indicating that non-target devices were resilient to the non-coincident pulses. These results prove the potential of our novel ReRAM technology for in-memory deep learning accelerators with fully parallel weight updates.

\subsection{In-Memory Neural Network Training Simulations}
\quad Based on the experimental results, we assess the in-memory NN learning performance by using the conventional stochastic gradient descent (SGD) algorithm. We use a NN structure of three layers with 784, 256, and 10 neurons consisting of 203,264 synaptic analog memories (\textbf{Figure 6a}). In the simulation, the soft-bounds model is used to emulate the realistic conductance update behaviors of analog ReRAM devices, as showcased by the colored lines in \textbf{Figure 6b}. The blue line represents our device model with $k=0.005$, while the black line displays an exemplary device model that is susceptible to the disturbing pulses ($k=0.2$). The red data represents the ideal case with $k$ = 0, as a reference for comparison. The blue line properly reproduces the electrical behavior of our device compared to the experimental data, as indicated by the gray open symbols. The parameters used for the model are $w_{\rm max}$ 1, $w_{\rm min}$ -1, $dw_{min}$ 0.6, $\sigma_{c\_to\_c}$ 0.5, and $\sigma_\pm$ -0.1. The details of the model can be refereed to the experimental in Section Methods\textsuperscript{\cite{ rasch2024NatureComm}}. We further validated the developed soft-bounds model by reproducing the behavior of multiple devices. The calculated $k$ distribution from the reproduced data corresponded well with the experimentally obtained results shown in Figure 5d (note that these results are not included).
In the simulator, not only the in-memory forward and backward propagation is implemented, but we also implement the parallel in-memory weight update scheme by generating stochastic bitstreams for the array inputs. The validity of the stochastic parallel update module can be confirmed by Figure S12 that shows statistical correlations between the desired number of update pulses of a weight matrix and the resulting number of coincident pulses of a ReRAM array. The setup simulates the fully in-memory training process by generating probability-encoded pulse trains, calculating coincident pulses of every cross-point element, and changing synaptic weights based on the established soft-bounds model.
\begin{figure}[!t]
  \includegraphics[width=\linewidth]{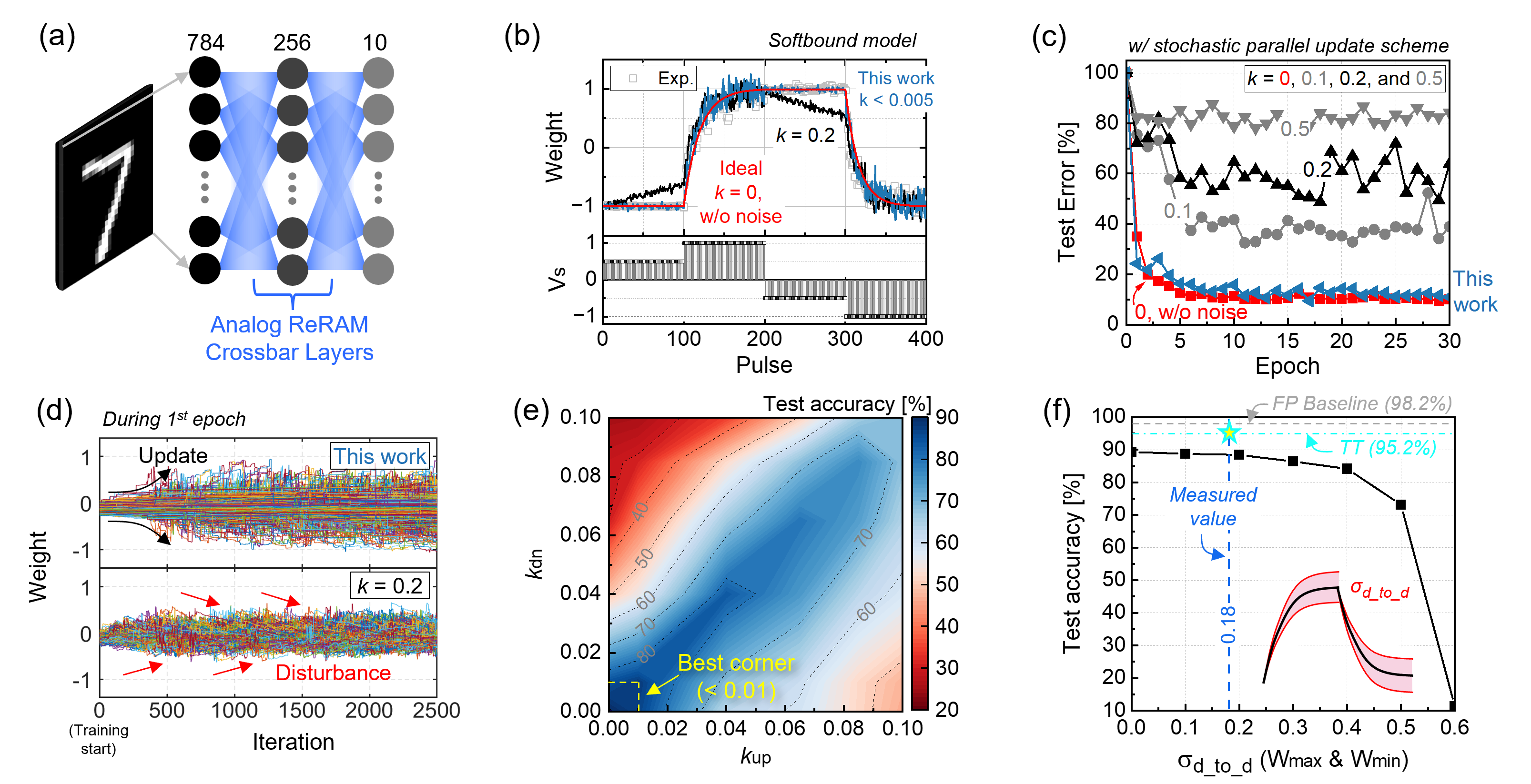}
    \caption{
    (a) Neural network structure for in-memory training simulation. The analog ReRAM crossbar layers and the stochastic parallel weight update scheme are considered in the simulation.
    (b) Simulated pulse responses of CMO/HfO${\rm x}$ ReRAM with $k$ = 0.005 (blue, this work) and $k$ = 0.2 (black) through the established soft-bounds models. The red data represents the ideal case with $k$ = 0, as a reference for comparison.
    (c) Test accuracies during in-memory training with different $k$ values, where the models in (b) are included. The NN hyperparameters for the training are a learning rate of 0.6 and 10k training images with a batch size of unity.
    (d) History of the synaptic weights during 2500 training iterations in the 1$^{\rm st}$ epoch.
    (e) Contour plot showing the test accuracy as a function of $k_{\rm up}$ and $k_{\rm dn}$ values. For successful learning convergence, $k$ less than 0.01 is desired as noted.
    (f) Test accuracy after training as a function of device-to-device noise ($\sigma_{d\_to\_d}$). The measured value of 0.18 is indicated by the blue dashed vertical line. The integration with the state-of-the-art Tiki-Taka (TT) algorithm enables achieving an accuracy of 95.2\% (bright blue), close to the floating-point (FP) baseline (gray).}
  \label{fig:F6}
\end{figure}

\quad The results in \textbf{Figure 6c} showcase test errors with different $k$ values as a function of training epoch. Here, symmetric $k$ ($k_{\rm up}$ = $k_{\rm dn}$) was assumed. The ideal case (red line) smoothly converges to the system’s optimum performance, achieving a test accuracy of 90.6\%. In contrast, the result of our case (blue line) shows slight fluctuations in test error but still converges successfully, achieving 89.08\% test accuracy. We can also observe the test accuracy of the black line with $k=0.2$ shows 48.82\% low performance. The simulations were conducted by using 10k training images and 10k unseen test images from the hand-written MNIST dataset. The $BL$ is dynamically adapted from 10 to 1 for better training efficiency\textsuperscript{\cite{ rasch2024NatureComm}}. 
The previous study reported that a 0.1 $k$ shows floating point (FP) baseline comparable learning performance with the CMOS-based linear weight update devices\textsuperscript{\cite{gokmen2016acceleration}}. In contrast, our results reveal that a 0.1 $k$ value  hinders the NN learning of the non-linear ReRAM-based hardware, increasing test errors by more than 20\%. This confirms that ensuring update disturbance-resilience is crucial for successful on-chip learning using analog emerging memories. The results also imply that the investigation of update disturbance effect must be conducted while considering realistic device characteristics, such as asymmetric update behaviors.
\textbf{Figure 6d} shows the weights evolution during 2500 training iterations from the initialized weights around the symmetry point. The synaptic weights of our ReRAM devices start to update their memory states properly, while not being affected by the non-coincident pulses during stochastic parallel array updates. With the presence of update disturbances, however, the non-coincident pulses consistently distort the updated weight states towards their preferable SP of the material (as shown in Figure 6d, the case of $k=0.2$). 

\quad Furthermore, we have conducted a systematic analysis on the learning performance with specified $k$ for up and down directions. \textbf{Figure 6e} displays a contour plot of final test accuracy across the $k_{\rm up}$ and $k_{\rm dn}$ values from 0 to 0.1. For successful convergence of the NN training, an ultra-low $k$ value, i.e. less than 0.01, is required as noted by the yellow dashed line in Figure 6e. Intriguingly, the results also show that the asymmetry in $k$ remarkably deteriorates the learning performance. The asymmetric responses may lead to one-sided weight shifting consistently, disrupting the NN learning. The potential mitigation methods will be further discussed in the discussion section.
Note that the learning efficiency with non-zero $k$ values can be highly affected by the optimization of hyperparameters, such as the size of training dataset, stochastic BL, and weight initialization. For the Figure 6e, the highest test accuracy was used with hyperparameter optimization. Overall, the $k$ value close to 0 is highly desired for on-chip learning convergence. Considering the integration with a state-of-the-art analog training algorithm, e.g., Tiki-Taka (TT), securing the update-disturbance resilience of analog memory devices is critical to operate the system correctly.
Since device-to-device variability, $\sigma_{d\_to\_d}$, is inherent in real hardware, we further investigate the impact of $\sigma_{d\_to\_d}$ noise in the $w_{\rm max}$ and $w_{\rm min}$ on the learning performance. \textbf{Figure 6f} reveals the test accuracy with respect to device-to-device noise magnitude in the conductance boundaries. As indicated by the vertical blue dashed line, the measured $\sigma_{d\_to\_d}$ of our device (0.18 from Figure 2f) represents a negligible penalty in the training accuracy. Based on our established ReRAM technology, the adoption of TT algorithm enhances the learning performance to 95.2\%, as highlighted by the horizontal bright blue line. For the training simulation, the TT algorithm version 4 was used by including the $\sigma_{d\_to\_d}$ noise. Together, our disturbance-robust ReRAM technology proves its great potential for in-memory AI hardware accelerator, achieving high training accuracies. 

\section{Discussion}
\begin{table}[ht!]
 \caption{Benchmark table of analog emerging memories for deep learning accelerator applications}
  \begin{tabular}{@{}|p{2.8cm}|p{2.1cm}|p{2.3cm}|p{2.3cm}|p{2.3cm}|p{2.0cm}|p{2.3cm}|@{}}
    \hline
        &  2023 Adv. Mater.  \cite{wu2023AM} &  2024 Trans. Elec. Dev.  \cite{kim2024ud} & 2024 Adv. Elec. Mat. \cite{chung2024ud} & 2023 Nat. Comm.  \cite{chen2023NatureComm} & 2024 IEDM \cite{Son2024IEDM.ECRAM} &  \textbf{This work} \\
    \hline
        Device type & ReRAM  & ReRAM  & ReRAM & ECRAM & ECRAM & ReRAM \\
    \hline
        Materials & \textcolor{red}{-} & Al$_2$O$_3$/HfO$_2$ & ZnO/NiO$_{\rm x}$/ ZnO & YSZ/WO$_{\rm x}$ & WO$_x$/HfO$_2$/ WO$_x$& CMO/HfO$_{\rm x}$ \\
    \hline
        Device diameter & 200 nm &  \textcolor{red}{50 \textmu m}  & \textcolor{red}{100 \textmu m} & \textcolor{red}{100 \textmu m} & 300 nm & 200 nm \\
    \hline
        Pulse width & \textcolor{red}{-} & 5 \textmu s & \textcolor{red}{640 \textmu s} & \textcolor{red}{100 ms} & \textcolor{red}{100 ms}& 2.5 \textmu s  \\
    \hline
        Pulse amplitude & +3/-3 V  & \textcolor{red}{+3/-5 V} & \textcolor{red}{+5/-4 V} & \textcolor{red}{+5/-5 V} & {+2/-1.5 V}&  {+1.4/-1.8 V} \\
    \hline
        $g_{\rm min}$ & 4 \textmu S  & 60 \textmu S &  10 nS  & 100 nS & 0.5 nS & 5 \textmu S   \\
    \hline
        $g_{\rm max}$ & 40 \textmu S & 120 \textmu S  & 140 nS  & 36 \textmu S & 7 nS & {60 \textmu S}   \\
    \hline
        Pulse cycling endurance & \textcolor{red}{-} & \textcolor{red}{-} & \textcolor{red}{-}  & 50 M & 10 M &100 M  \\
    \hline
        $k$ & \textcolor{red}{-} & \textcolor{red}{0.1} & \textcolor{red}{0.13} & \textcolor{red}{0.03} & \textcolor{red}{0.03} & $<0.005$ ($^*$std: 0.025) \\
    \hline
        Tested \# of pulses & 
        - & 
        30 & 
        100 & 
        10 & 
        100 & 
        100 k \\
    \hline
        Tested devices & 
        - & 
        1 & 
        1 & 
        4 & 
        1 & 
        15 \\
    \hline
        Tested states & 
        - & 
        G$_{\rm min}$ & 
        G$_{\rm min}$ & 
        G$_{\rm min}$ & 
        G$_{\rm min}$ & 
        G$_{\rm min}$ \& G$_{\rm max}$ \\
    \hline
        BEOL integration (technology) & 65 nm &  -  & \textcolor{red}{-} & \textcolor{red}{-} & \textcolor{red}{-} & 0.35 \textmu m \\
    
    \hline
  \end{tabular}
      \small
      $^*$std: standard deviation
  \label{T1}
\end{table}

\quad \textbf{Table 1} provides an overview of various emerging memory devices including this work, highlighting their key specifications. Our analog ReRAM technology exhibits competitive performance in switching operations, while demonstrating superior update-disturbance robustness compared to other technologies. Wu et al. reported an areal uniform bulk-switching type ReRAM device for deep learning accelerators\cite{wu2023AM}. However, the update disturbance analysis is lacking. 
We also discuss potential device candidates for a disturbance-tolerant analog memory device. One solution is $IV$ rectifying ReRAM devices with a highly insulating barrier layer, inspired by the selector-less ReRAM for conventional memory applications (Table \ref{T1})\textsuperscript{\cite{kim2024ud,kim2023retention,chung2024ud}}.  
However, this does not always guarantee low $k$ value for the device, as shown in the Table 1.
This raises an important point: the definition of device update disturbance for analog in-memory training applications needs to be reconsidered. Also, the analog memory device should be thoroughly studied by conducting proper investigation methods.
{Moreover, the insulating barrier layer of $IV$ rectifying ReRAM significantly contributes to the voltage drop across the device material stack, requiring high voltage and long pulses to initiate device programming. These properties are undesirable for deep learning acceleration and seamless integration with advanced CMOS nodes\textsuperscript{\cite{gong2022iedm, kim2021edl.rram}}.} 
Another emerging memory candidate for AI training accelerators is the electrochemical transistor, known as ECRAM\textsuperscript{\cite{lee2021VLSI.ECRAM,lee2021Nanotech.ECRAM,lee2020EDL.ECRAM,Lee2021IEDM.ECRAM}}. As shown in Table \ref{T1}, a recent study has reported its exceptional performance, including highly granular memory states, low switching variability, and low power operation\textsuperscript{\cite{cui2023NatureElec.ECRAM, Son2024IEDM.ECRAM}}. However, the millisecond-long pulse width and a $k$ value as large as 3\% must be further improved for successful deep learning acceleration. Moreover, the large device area needs further optimization for a highly scalable hardware system.
In contrast, the proposed CMO/HfO$_{\rm x}$ analog ReRAM exploits the pre-formed nanoscale CF to effectively localize the analog resistive switching, which enhances the switching non-linearity. \textbf{Table 1} demonstrates the favorable behavior of our devices, which do not exhibit a notable G shift in the direction of disturbance even after 100k pulses, resulting in $k$ < 0.005 with 0.025 standard deviation. In contrast, other studies showed significant G drifts after only 100 pulses, or even less. A higher number of non-coincident pulses is desired for disturbance tests. Furthermore, systematic investigations at both G$_{\rm min}$ \& G$_{\rm max}$ across multiple devices is essential for the reliability evaluation.

\quad In Figure \ref{fig:F6}e, we showed that the asymmetry in the non-linearity factors between $k_{\rm up}$ and $k_{\rm dn}$ can cause the substantial accuracy loss. Here, we also briefly discuss potential mitigation methods for optimizing deep learning performance with the asymmetry in the non-linearity factor. First, the pulse scheme can be modified when designing the baseline voltage level of stochastic bitstreams for the array inputs. The conventional pulse design is commonly based on the half-bias scheme with a same baseline voltage level so that the effective potential difference at the baselines constitutes zero (non-selected case). However, the half-selection pulse amplitude is reduced by setting the baseline voltage differently for rows and columns, such that the baseline potential difference shifts to a non-zero value towards the opposite polarity of the half-selection pulses. Thereby, the voltage magnitude at the half-pulses can be reduced at the cost of non-zero baseline voltage. Thus, adjusting the baseline voltage can help balance the asymmetric $k$, alleviating accuracy loss. Additionally, further optimization can be made by modifying hyperparameters for NN learning, such as increasing the probability of stochastic pulses and simultaneously reducing the bitstream length $BL$. This helps minimize the frequency of non-coincident pulses, while maintaining the frequency of coincident pulses. Also, a large number of training iterations increases the absolute number of non-coincident pulses, which can potentially amplify disturbance effects. As a result, the size of the training dataset should be optimized in the presence of update disturbance. 

\section{Conclusion}
We demonstrated a disturbance-resilient analog ReRAM on 350 nm silicon node with CMOS compatible materials and processes. The device also features compelling properties for in-memory training accelerators, e.g., 60 ns fast linear switching and analog memory states during open-loop operation. We identified the significant role of the nanoscale filament in inducing thermoelectric energy concentration during resistive switching through COMSOL Multiphysics simulations. The high non-linearity in the switching responses to the input voltage was extensively investigated. Furthermore, disturbance-tolerant, parallel weight mapping was experimentally demonstrated on a ReRAM array chip. Finally, the hardware-aware NN simulations presented comprehensive assessments of the learning accuracy, by considering the experimental findings. The results highlight great promise of our CMO/HfO$_{\rm x}$ ReRAM technology for in-memory training solutions.

\section{Experimental Section}
\threesubsection{Chip Fabrication}\\
\quad The ReRAM material layers are integrated with a 0.35 \textmu m CMOS chip at Back-End-Of-Line. On the intermediate dielectric layer, a 20 nm TiN bottom electrode and a 4 nm sub-stoichiometric HfO$_{\rm x}$ layer are deposited using plasma-enhanced atomic layer deposition (PEALD) at 300 \textdegree C, as indicated in Figure \ref{fig:F2}b. After forming a 20 nm conductive metal oxide layer, a 20 nm TiN top electrode is deposited by sputtering, followed by a 50 nm W capping layer. The active ReRAM structure is defined using a dry etching tool, and the surface is passivated with a Si$_3$N$_4$ layer using PECVD. After opening the via to access the cell, the final W metal layer is deposited for routing. The fabricated device dimensions range from 200 nm to 2 \textmu m. Further investigations on scaled devices were also conducted and confirmed down to 70 nm on a silicon dioxide wafer. The sizes of the BEOL-integrated ReRAM arrays range from $2 \times 2$ to $10 \times 10$ with 5 \textmu m electrode line width.\\

\threesubsection{Electrical Characterization}\\
\quad For the fast pulse measurements in the sub-microsecond regime, data were collected using a Keithley 4200A machine and pulse measurement units (PMU). The array measurement setup was built based on a host computer and a National Instruments (NI). The setup is controlled by a Python-based graphical user interface (GUI) on the host computer. The multichannel analog output (PXIe-6739), analog input (PXIe-4309), and switch (PXIe-2571) modules in the NI chassis are interconnected through custom-designed PCBs. The pulses generated by the analog output module on each channel are applied to the array rows and columns, passing through 100 $\ohm$ sense resistors connected to the analog input module for sensing. This configuration enables the setup to perform both biasing and reading simultaneously. Note that an additional switch module is used to interface the array with the analog I/O modules, allowing access only during measurements. For reading, a 0.2 V pulse was used with 100 kHz sampling rate. 

\threesubsection{Neural Network Simulation}\\
\quad The hardware-aware NN simulations for on-chip training were conducted by using MATLAB and the extended simulations with Tiki-Taka algorithm were performed using IBM AI hardware toolkit (AIHwKit)\textsuperscript{\cite{rasch2021AICAS,le2023aihwkit}}. For implementing synapses with negative weights, we assumed the use of a subtraction scheme with main analog memory arrays\textsuperscript{\cite{rasch2024NatureComm,haensch2018IEEEproceedings}}. The reference resistors were pre-programmed to the middle value of the device conductance range, which corresponds to $w=0$. In the experiments, the sigmoid function was used for neuron activations, except for the last-layer neurons, which used the softmax function. The simulations were carried out using conventional stochastic gradient descent (SGD) algorithm for a generalized analysis. The fully connected NN structure with 784, 256, and 10 neurons was adopted, containing 203,264 analog ReRAM memories. The training begins after initializing the devices to the symmetry point (SP), by applying one-up and one-down pulse set repetitively\textsuperscript{\cite{ rasch2024NatureComm, kim2019zero}}. 
The stochastic pulse-based in-memory training module and device material model implemented in the simulator are described in following subsection.\\

\threesubsection{In-Memory Outer-Product Weight Update}\\
\quad 
From the perspective of a synapse connected to i$_\text{th}$ pre-neuron and j$_\text{th}$ post-neuron, the desired local weight update is defined as 
\begin{equation} \label{E2}
w_{ij} \leftarrow w_{ij} + \eta x_i d_j
\end{equation}
where $\eta$ represents a pre-defined learning rate.  To simulate the stochastic fully parallel weight update scheme, a stochastic translator is implemented to generate a probabilistic bit, $P$, based on the corresponding neuron signals (e.g., $x_i$ and $d_j$). When submitting the stochastic bitstreams to an array, the total number of coinciding pulses that a cross-point synapse experiences can be expressed by
\begin{equation} \label{E3}
N_{\text{update}\_{ij}} = \sum_{n=1}^{BL} P_{i}^n \wedge P_{j}^n
\end{equation}

where $BL$ represents stochastic bitstream length. As a result, $N_{\text{update}\_{ij}}$ becomes proportional to the desired weight update, $\eta x_i d_j$, as shown in Equation \ref{E3}. Similarly, a non-coinciding pulse case is estimated by $P_i^n \vee P_j^n$ at the n$_{\rm th}$ bit of the bitstream. As the simulation accounts for realistic situations where the cross-point devices receive either full or half pulses sequentially throughout the entire bitstream, the in-memory training module repeats the AND and OR operations for every cross-point element until the end of $BL$. Based on these results, the incorporated device material model implements realistic, nonlinear weight changes per pulse at the same time. Note that BL is initially set to $\eta / dw_{\rm min}$ in order to match the learning rate in the FP baseline model\textsuperscript{\cite{gokmen2016acceleration}}.\\

\threesubsection{Device Material Model}\\
\quad We adopted the soft-bounds model as described in our previous study\textsuperscript{\cite{rasch2024NatureComm}}. The weight change $\Delta w$ to a programming pulse is given by 
\begin{equation} \label{E4}
\begin{split}  
\Delta w^+ = \alpha^+ {(\frac{\Breve{w}_{\rm max} - w}{\Breve{w}_{\rm max}}+\sigma_{\rm c\_to\_c} \xi)} \\
\Delta w^- = -\alpha^- {(\frac{\Breve{w}_{\rm min} - w}{\Breve{w}_{\rm min}}+\sigma_{\rm c\_to\_c} \xi)}
\end{split}
\end{equation}

where $\alpha^+$ and $\alpha^-$ represent the slope parameter in each direction, and $\sigma_{\rm c\_to\_c}$ corresponds to cycle-to-cycle update fluctuations. $w$ corresponds to the current weight state of the device. As more pulses are applied to the device, $\Delta w$ gradually decreases toward zero, approaching the conductance bounds, i.e., $\Breve{w}_{\rm max}$ and $\Breve{w}_{\rm min}$. Here, we also introduced device-to-device noise at the saturated boundaries: $\Breve{w}_{\rm max} = w_{\rm max} (1 + \sigma_{\rm d\_to\_d} \xi)$ and $\Breve{w}_{\rm min} = w_{\rm min} (1 + \sigma_{\rm d\_to\_d} \xi)$, where $\xi \in \mathcal{N}(0,1)$. As we consider device-to-device variations only in the boundaries for the moment, the slope $\alpha$ can be simplified as $\alpha^+ = dw_{\rm min}(1 + \sigma_\pm)$ and $\alpha^- = dw_{\rm min}(1 - \sigma_\pm)$ for each direction. The $\sigma_\pm$ provides the slope difference between the up and down directions, which affects the position of the symmetry point (SP). The $dw_{\rm min}$ is a material parameter representing the average update response to a single pulse at the SP.
The final amount of weight change considering disturbance, $\sigma_{\rm k}$, is defined as
\begin{equation} \label{E5}
\begin{split}
    \Delta\Breve{w}^+ = 
    \begin{cases}
        \Delta w^+, & \text{if } V_s=1 \\
        \Delta w^+ \sigma_k^+, & \text{if } V_s=0.5
    \end{cases}
\\
    \Delta\Breve{w}^- = 
    \begin{cases}
        \Delta w^-, & \text{if } V_s=1 \\
        \Delta w^- \sigma_k^-, & \text{if } V_s=0.5
    \end{cases}
\end{split}
\end{equation}
where $V_s$ represents the normalized switching voltage of applied pulses. The $V_s = 1$ case corresponds to when the device receives a coincident pulse, which initiates a proper weight change. On the other hand, the $V_s = 0.5$ case corresponds to when the device receives a non-coincident half pulse, which causes an undesired weight change. The resulting weight update per pulse of a synapse is described as
\begin{equation} \label{E6}
w \leftarrow w +     \Delta\Breve{w} 
\end{equation}
Note that $\sigma_{k}$ reflects the update disturbance due to the non-coincident pulse and cannot be directly derived from the non-linearity factor, $k$. Therefore, a lookup table method is adopted to correlate with the $k$ of the devices. It is also worth noting that additive random noise is introduced to the updated $w$ to reproduce realistic device responses, with standard deviations of 0.1 for coincident switching and of 0.025 for non-coincident cases, respectively. This disturbance-incorporated soft-bounds model is beneficial for reproducing realistic pulse responses, considering the current $w$ state of the devices.\\

\medskip
\textbf{Supporting Information} \par 
Supporting Information is available from the Wiley Online Library.

\medskip
\textbf{Acknowledgements} \par 
This work is co-funded by SNSF ALMOND (grantID: 198612), by the European Union and Swiss state secretariat SERI within the H2020 MeM-Scales (grantID: 871371) and PHASTRAC (grantID: 101092096) projects. The authors also acknowledge the Binnig and Rohrer Nanotechnology Center (BRNC) at IBM Research Europe - Zurich.

\medskip
\bibfont{\small}
\bibliographystyle{MSP}
\bibliography{REF}



\end{document}